\documentclass[]{emulateapj}

\slugcomment{Submitted to The Astrophysical Journal}

\shorttitle{Supermassive Black Hole Binary Evolution}
\shortauthors{Khan et al.}

\begin{document}


\title{Formation and Hardening of Supermassive Black Hole Binaries in Minor Mergers of Disk Galaxies}


\author{Fazeel Mahmood Khan\altaffilmark{1,2}}
\author{Ingo Berentzen\altaffilmark{1,3}}
\author{Peter Berczik\altaffilmark{4,1,5}}
\author{Andreas Just\altaffilmark{1}}
\author{Lucio Mayer\altaffilmark{6}}
\author{Keigo Nitadori\altaffilmark{7}}
\author{Simone Callegari\altaffilmark{6}}
\affil{$^1$Astronomisches Rechen-Institut, Zentrum f\"ur Astronomie, 
Univ. of Heidelberg, M{\"o}nchhofstrasse 12-14, 69120 Heidelberg, Germany}
\affil{$^2$Department of Physics, Government College University (GCU), 54000  
Lahore, Pakistan}
\affil{$^3$Heidelberg Institute for Theoretical Studies, Schlosswolfsbrunnenweg 35, 69118 Heidelberg, Germany}
\affil{$^4$National Astronomical Observatories of China, Chinese Academy of Sciences, 20A Datun 
Rd., Chaoyang District, 100012, Beijing, China}
\affil{$^5$Main Astronomical Observatory, National Academy of Sciences of Ukraine, 27 Akademika 
Zabolotnoho St., 03680, Kyiv, Ukraine}
\affil{$^6$Institute for Theoretical Physics, University of Z\"urich, Winterthurerstrassse 190, CH-9057 Z\"urich, Switzerland }
\affil{$^7$RIKEN Institute, Tokyo, Japan}

\begin{abstract}
%%At a separation of approximately one parsec, binary supermassive black holes (SMBHs) stall in spherial galaxy model. In an earlier %%study, we show that stalling does not happen for equal mass galaxy mergers. Galaxies that form via major mergers are substantially %%nonspherical, and it has been argued that the centrophilic orbits in triaxial galaxies might provide stars to the massive binary at a %%high enough rate to avoid stalling. Here using a set of direct N-body simulations of unequal mass mergers with different galaxy(and %%SMBH) mass ratios for different density profiles of merging galaxies, we study the shape of merger remnat and the SMBH binary %%hardening rates, we find .................
We model for the first time the complete orbital evolution of a pair of Supermassive Black Holes (SMBHs) in a 1:10 galaxy merger of two disk dominated gas-rich galaxies, 
from the stage prior to the formation of the binary up to the onset of gravitational wave emission when the binary separation has shrunk to 1 milli parsec. 
The high-resolution smoothed particle hydrodynamics (SPH) simulations used for the first phase
of the evolution include star formation, accretion onto the SMBHs as well as feedback from supernovae explosions and radiative heating from the SMBHs themselves.
Using the direct $N$-body code $\phi$-GPU we evolve the system further without including the effect of gas, which has been mostly consumed by star
formation in the meantime. We start at the time when the separation between two SMBHs is $\sim 700$ pc and the two black holes are still embedded in their
galaxy cusps. We use 3 million particles to study the formation and evolution of the SMBH binary till it becomes hard. After a hard binary is formed, we reduce (reselect) the particles to 1.15 million and follow the subsequent shrinking of the SMBH binary due to 3-body encounters with the stars.
We find approximately constant hardening rates and that the SMBH binary rapidly develops a  high eccentricity.
Similar hardening rates and eccentricity values are reported in earlier studies of SMBH binary evolution in the merging of dissipationless
spherical galaxy models. The estimated coalescence time is $\sim 2.9$~Gyr, significantly smaller than a Hubble time. We discuss why this timescale
should be regarded as an upper limit. Since 1:10 mergers are among the most common interaction events for galaxies
at all cosmic epochs, we argue that several SMBH binaries should be detected with currently planned space-borne gravitational wave interferometers, whose             
sensitivity will be especially high for SMBHs in the mass range considered here.
%, have grown to, $1.5 \times 10^6 \mathrm{M}_{\odot}$ and $4.8 \times 10^5 \mathrm{M}_{\odot}$ respectively

\end{abstract}



\section{Introduction}\label{sec-intro}

 Central supermassive black holes (SMBHs) are ubiquitous and are found in a variety of galaxies, ranging 
 from low mass galaxies to the most massive early-type galaxies \citep{FF05,gul09}. Within our current  
 cosmological picture of hierarchical structure formation, galaxies form through continuous mergers.
 If both candidate galaxies harbor a central SMBH before the merger, the evolution of the latter is thought to be as follows  \citep{beg80}: the SMBHs
 of the merging galaxies sink towards the center of the merger remnant due to dynamical friction and form 
 a gravitationally bound binary system. The further evolution of the SMBH binary is governed by interactions 
 with stars and gas. If the binary semi-major axis value shrinks to a value where emission of gravitational waves (GWs) 
 efficiently takes away energy and angular momentum from the binary, the coalescence of SMBHs becomes inevitable. 
 In fact there is growing observational evidence for this process: there are reports about two widely separated SMBHs in a single 
 galaxy \citep[e.g.][]{kom03,rod06,fab11} as well as indirect evidence for binary 
 SMBHs \citep[e.g.][]{mer02,liu03,val08,bor09,igu10}. 

 Binary SMBHs are of particular interest in astrophysics and general relativity: if  SMBH binary evolution
 leads to coalescence following a galaxy merger, such an event would give rise to one of the {\it loudest} 
 possible bursts of GWs detectable for future space borne low-frequency laser interferometers
 \citep{Hughes03,BC04,LISA2012}. However, the exact pathway to SMBH coalescence and especially the
 associated timescales, and hence the chances for a possible detection during any satellite mission
 run-time, is still a matter of debate.

Numerical $N$-body simulations considering the decay of a pair of SMBHs in major mergers of massive
elliptical galaxies due to the interaction with the stellar background only show that the critical separation 
for gravitational wave emission should be reached in about one Gyr as long as the galaxies deviate sufficiently
from sphericity, a configuration which allows to have centrophilic orbits \citep{st99,pm01,MP04,mv11} that efficiently refill
the loss cone \citep{ber06,ber09,kha11,kha12,pre11}. \citet{kha12} showed that SMBHs with masses ranging
from $10^6 \mathrm{M}_{\odot} - 10^7 \mathrm{M}_{\odot}$, e.g., at the center of faint elliptical galaxies or
in the bulge of spiral galaxies, coalesce well within a Gyr after the merger of the two galaxies/bulges. However,
these simulations started from idealized initial conditions and do not consider the effects of gas dynamics,
star formation, etc.  

Conversely, numerical simulations investigating major mergers including hydrodynamics show that a SMBH
binary can form very rapidly after the galaxy collision takes place, sometimes in less than a million
years \citep{may07}, but are not yet conclusive on the subsequent shrinking of the binary at sub-parsec
scale separations due to the prohibitive cost of the computation with increasing resolution and
the uncertainties in the modeling of gas thermodynamics and turbulence \citep{cha11}. 

Furthermore, minor mergers are the most frequent type of mergers in the Universe, especially those
 with a mass ratio of some $1:10$.  In addition, they usually involve disk-dominated galaxies because these
 are not only the most common galaxies  in the Universe today \citep{bin88} but they
were even more common at higher redshift since they are the likely progenitors of present-day massive early type
 galaxies \citep{van11}.

Simulations of minor mergers between disk galaxies show that the role of gas can be
crucial in delivering the pair of SMBHs to the center of the merger remnant; without gas the SMBH of the secondary galaxy is left wandering at kilo-parsec distances, at which the dynamical friction timescale is longer than
the Hubble time, because the core of the galaxy that is hosting it is tidally disrupted before it can sink
to the center of the primary \citep{cal09}. With the gas the pairing of the two SMBHs is successful because the gas-rich merging satellite undergoes
a strong central star formation burst, developing a higher central density allowing its survival to tidal disruption down
to the center of the primary. \citet{cal11}, who studied the pairing of SMBHs in the merger of late type gas rich galaxies in simulations which include star formation
and accretion onto the SMBH, found that the mass ratio of the two SMBHs can change significantly due to the fact that the secondary is fed at a higher rate as a result of the stronger gas inflow caused by the tidal disturbance. The latter simulations of minor mergers could not follow the decay of the two SMBHs to separations of less than tens of parsec. Hence it is unknown whether or not they would reach the critical separation for gravitational wave emission and on which timescales. Continuing the calculations further with gas dynamics is computationally challenging. In addition, at the end of the merger significant star formation takes place, increasing the contribution of the stars to the potential relative to the gas.

Therefore it is sensible to consider the approximation in which the stars are the dominant source of the drag acting onto the SMBHs during the last stage of the sinking before gravitational waves take over. This is what we attempt to do for the first time in  this paper, starting from a realistic late configuration of a gas-rich minor merger performed by \cite{cal11}, we followed the evolution of the SMBH binary to the milli parsec (mpc) regime where GWs start becoming important.  

The paper is organized in the following way: In Section~2 we describe the initial conditions for our direct $N$-body simulations. The numerical methods and codes used for our simulations are described in Section~3. In Section~4 we explain the results for the evolution of the SMBH binary. Finally, Section~5 concludes the findings of our study.  

\section{Initial Conditions}\label{sec-model}

\begin{figure}
\centering
\includegraphics[width=0.99\columnwidth]{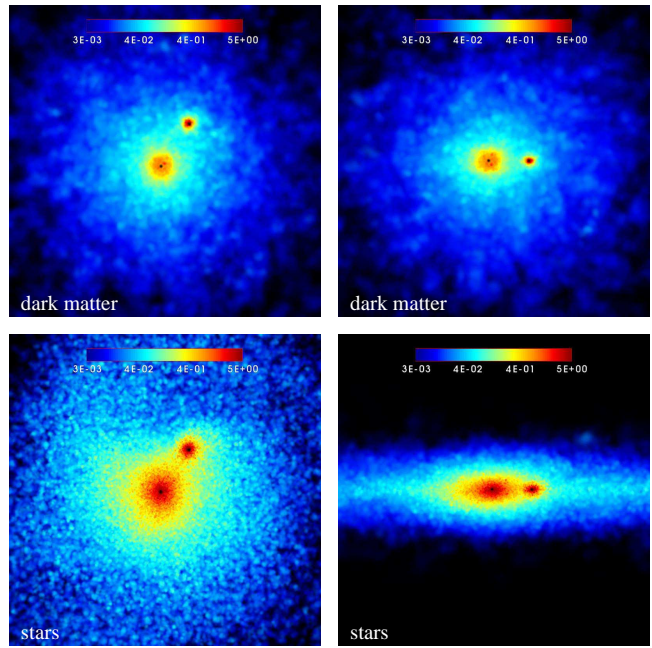}
\caption[]{
 Density distribution of the dark matter (top panels) and stellar component (bottom panels) in
 the $x-y$ (left column) and $y-z$ (right column) planes. The two high density regions are
 clearly visible around the two SMBHs (black dots) in the center. The size of each box is 4 kpc.
} \label{fig1zu}
\end{figure}

Owing to numerical limitations in simulations of galaxy mergers in respect of accuracy on parsec or sub-parsec scale, e.g., due to approximations in the force calculation (gravitational softening, tree-approximation) or insufficient integration 
schemes we follow a multi-step strategy to  study the evolution of binary SMBHs. We take 
simulation data from a galaxy merger simulation including SMBHs  in its final phase, i.e., before reaching resolution
 limits and then follow up these simulations in high-resolution using direct $N$-body calculations with the dynamically relevant though spatially truncated region of the original galaxy merger.
 
\cite{cal11} studied the pairing of binary SMBHs in several 1:10 mergers of two disk galaxies using
 $N$-body/Smoothed Particle Hydrodynamics (SPH) simulations conducted with the GASOLINE code \citep{wad04}.
The reference galaxy model in that study was a Milky Way type disk galaxy consisting
of three components: (i) a spherical and isotropic Navarro, Frenk \& White profile dark matter halo, 
(ii) an exponential disk consisting of stars and gas and (iii) a spherical Hernquist bulge.
 The mergers considered different orbital configurations. In all cases, except for a peculiar retrograde merger case, the two SMBHs paired at the
center of the remnant in less than 1 Gyr from the end of the galaxy collision.
In this paper we employ one particular merger set-up as initial conditions since already a single simulation
of this kind requires several months of computation. In particular we choose the coplanar pro-grade merger simulation with 30\% gas fraction in the disk
since it resulted in the shortest merging time for the galaxies.

The gravitational softening adopted in the study of \cite{cal11} was $\epsilon = 45$ pc for dark matter and baryonic particles in the larger galaxy, whereas for smaller galaxy it was 20 pc. The SMBH was introduced at the center of each galaxy represented by a point mass particle. The adopted masses for the SMBHs were $6 \times 10^5 \mathrm{M}_\odot$ and $6 \times 10^4 \mathrm{M}_\odot$  for the primary and satellite galaxy respectively, consistent with the $M_{\bullet}$-$M_{bulge}$ relation \citep{mer01,har04}. The simulation also includes the effects of star formation and accretion on to the SMBHs, as well as feedback from both processes. The final separation of the two SMBHs at the end of the simulation was $\sim 30$ pc, which is comparable to the softening that was used in the simulations. For more details of the construction of the galaxy models and the set up for the initial orbit of the galaxy merger we refer the reader to \citet{cal11}. 
 Here it suffices to say that the merger was assumed to start at $z \sim 3-5$, and is completed at a time corresponding
to $z \sim 1-2$, a choice motivated by the optimal detection window of planned laser interferometers such as eLISA. The choice of galaxy types and sizes, as well as
of the masses of the SMBHs just reported, is thus made based on the characteristic densities expected in the concordance $\Lambda$CDM model for galaxies having
the same circular velocity as the Milky Way $V_c \sim 200$ km s$^{-1}$.

In this study we choose a snapshot of the system at the time 2.56 Gyr of \citet{cal11}, when the separation between the two black holes is $~700$~pc and the black holes are still embedded in the individual cusps (Figure~\ref{fig1zu}). The masses of SMBHs are $M_{\bullet1} = 1.5 \times 10^6 \mathrm{M}_{\odot}$ and $M_{\bullet2} = 4.9 \times 10^5 \mathrm{M}_{\odot}$ corresponding to a mass ratio $q = M_{\bullet 2}/M_{\bullet1} = 0.3$.
% due to the different rates at which the gas is funnelled towards the center of two galaxies and hence accreted by the SMBHs. 
We select all particles in the central 5 kpc region of the merger remnant and added all those particles, which have pericenter passage smaller than 3 kpc. 

\begin{figure}
\centerline{
\resizebox{0.99\hsize}{!}{\includegraphics[angle=270]{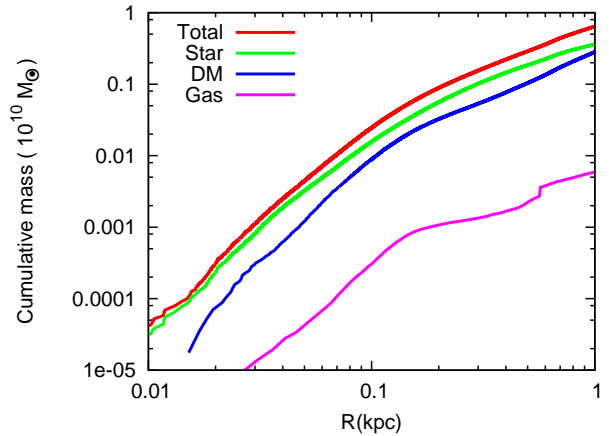}}
  }
\caption[]{
Cumulative mass profiles of the different components centered at the SMBH $M_{\bullet1}$ of the more massive galaxy at 
time 2.56 Gyr \citep{cal11}. This corresponds to the same time when we
select our particle sample for direct $N$-body simulations ($T=0$). 
} \label{fig2zu}
\end{figure}

Most of the gas in the central region is already converted into stars. The remaining gas particles, which contribute only a few percent in mass in the central region (Figure \ref{fig2zu}) are also treated as stars in our simulations. The total mass of the selected sample is $3.3 \times 10^{10} M_{\odot}$.

%\begin{figure}[!t]
%\vspace{-2cm}
%\centering

%\includegraphics[width=8.2cm, angle=0]{XY-ALL-den-0000.eps} \hfill
%\includegraphics[width=8.2cm, angle=0]{YZ-ALL-den-0000.eps} \\
%\includegraphics[width=8.2cm, angle=0]{XY-DM-den-0000.eps} \hfill
%\includegraphics[width=8.2cm, angle=0]{YZ-DM-den-0000.eps} \\
%\includegraphics[width=8.2cm, angle=0]{XY-STAR-den-0000.eps} \hfill
%\includegraphics[width=8.2cm, angle=0]{YZ-STAR-den-0000.eps} \\

%\caption{Spanshot at time = 0.0 Myr. top - all, midle - dm, bottom - stars.}

%\protect\label{fig1zu}
%\end{figure}

%\begin{figure}
%\centering
%\includegraphics[angle=270, width=0.75\columnwidth]{0005a.ps}
%\includegraphics[scale=1]{0022DZ.ps}
%\includegraphics[angle=270, width=0.75\columnwidth]{0005DZa.ps}
%\caption[]{
%Particles postions (inner 5 kpc) projected on initial orbital plane (top) and projected stellar densities (bottom). The two high density regions are clearly visible around the two %SMBHs (green points) in the center.
%} \label{fig1zu}
%\end{figure}

Although state-of-the-art hydrodynamical simulations, the Callegari et al. (2011) mergers use
a few million particles to represent the stellar component of a galaxy,
it is still several orders of magnitude smaller than the actual number of stars in a 
real galaxy. Because of the small number of dark matter particles the mass of one particle 
in the primary galaxy is comparable to the mass of the SMBH in the satellite galaxy
in \citet{cal11}. This is a potential problem for continuing the calculation to higher resolution in a regime
in which the  interaction with the background of stars and dark matter might dominate, since it can 
lead to unrealistic mass segregation of the dark matter and non-physically large kicks of the two SMBHs.
To avoid such high mass particle encounters with the SMBH binary, we split each dark matter particle in 
the primary galaxy into ten particles. Each child particle has mass 1/10 of the parent particle. 
The child particles are randomly distributed over a 10 pc sphere corresponding to the softening of 
the parent dark matter particle used in our simulations (see Section 3.2.). The child particles have the same velocities 
as their parent particles. A similar technique of particle splitting, which conserves linear momentum, was applied 
by \citet{may10} for SPH particles. We tested that split particles do not lead to artificial clumping.

We change the units of the original model from physical 
units to some model units. We choose our model units, so that the total mass of the galaxy ($3.3 \times 10^{10} M_{\odot}$)  $M_\mathrm{gal} = 1$, the length unit R is 1 kpc and also G = 1. This results in a time unit of $2.6$ Myr, velocity unit of 376 km s$^{-1}$ corresponding to a speed of light $c = 795$ in model units. 

For Run-1, which starts at $T=0$ (corresponding to the merger snapshot at time 2.56 Gyr of \citet{cal11}), we use $N = 3071296$ particles and evolve the system to the point, where the two cusps are merged and a ``hard binary" is formed. In order to increase the computational speed, we reduce the number of particle by selecting those, which have a pericenter passage inside 1 kpc $N = 1153984$ for Run-2. To resolve the stellar encounters with the SMBH binary at pericenter passage during the late phase of evolution, we reduce the softening of star-black hole interactions (see Section \ref{sec-nm}) in Run-3. Also for Run-3, we again split all dark matter particles having pericenter at less than 50 pc leading to $N = 1327488$. The orbit of the SMBH binary is integrated with higher accuracy during the late phase of the binary evolution in Run-4.

\section{Numerical Methods} \label{sec-nm}

\subsection{Simulation Software}

We use the direct $N$-body code $\varphi$-GPU with 4$^{\rm th}$ order Hermite integration
 scheme and hierarchical block time steps for our $N$-body simulations. 
%It has been developed from earlier 
%published code version $\varphi$GRAPE \citep{harfst} (which originally uses the {\tt 
%GRAPE6a} cards as a hardware accelerator for the calculations of the particles mutual 
%gravitational interaction).

The code is written from scratch in {\tt C++} and is  
based on an earlier {\tt C} version of $\varphi$-GRAPE\footnote{\tt 
ftp://ftp.ari.uni-heidelberg.de/staff/berczik/phi-GRAPE/}
designed for {\tt GRAPE6a} clusters \citep{har07}.
In the present version of the $\varphi$-GPU code we use native GPU support
and direct code access to the GPUs using only the NVIDIA
CUDA library\footnote{\tt http://www.nvidia.com/object/cuda\_home\_new.html}.
The multi-GPU support is achieved through MPI parallelization.
Each MPI process uses only a single GPU, but we usually start
two or more MPI processes per node (to effectively use the
multi core CPUs and the multi GPUs on our clusters).

The $\varphi$-GPU code is fully parallelized using the MPI library.
 The MPI parallelization
 was done in the same ``j'' particle parallelization scheme
  as in the earlier $\varphi$-GRAPE code. All the particles are
  divided equally between the working 
nodes (using the {\tt MPI\_Bcast()} command) and in each node we calculate only the 
fractional forces for the particles in the current time step, i.e. the so called
 ``active'' or ``i'' particles.
 We get the full forces from all the particles acting on the active particles after 
the global {\tt MPI\_Allreduce()} communication routine is applied.

 Besides the used 4$^{\rm th}$ order Hermite integration scheme, $\varphi$-GPU
 additionally supports a 6$^{\rm th}$ and even 8$^{\rm th}$ order Hermite
 integration. The numerical integration of the particle orbits as well
 as the time step criterion (see Section~3.3 for details) is based on the
 (serial CPU) $N$-body code YEBISU \citep{nk08}.
 
 More details about the $\varphi$-GPU code can be found in \citet{ber11}. 
 %\footnote{\tt http://www.hpc-ua.org/sites/default/files/proceedings-2011/1.1(8).pdf}. 

 The present version of the code used here is extensively modified
 to handle computational challenges required for our current project.

% In order to achieve full coalescence of he SMBH binary it is important to have zero softening for SMBHs. But the use of zero softening for the stars and dark matter leads to the %formation of binaries in the system which though is more realistic but causes the enormous slow down of simulations because of small time steps need to resolve the orbits of these %binaries. $\varphi$-GPU suports the use of different softenings for different components  . We use three different softenings ($\epsilon$): it is $\epsilon_{bh} = 0$ when calculating the %pairwise forces between the two SMBHs, $\epsilon_{s} = 0.001$ pc for star star interaction and $\epsilon_{dm} = 10$ pc for drak matter - drak matter interaction. For the interactions %betwen two different components, we adopt the mean of the two softenings. A softening of $\sim$ of 0.001 pc is necessary for the star-SMBH interaction
%to bring the SMBH binary separation to mili parsec where GW emission become efficient. 
% The time step criteria for for individual particles is optimized for the case in which the total mass of the system is $\sim$ unity. So we change the units of sytem from physical %unit to model units. In our model units, the total mass of the galaxy ($3.3 \times 10^{10} M_{\odot}$)  $M_{gal} = 1$, the length unit is 1 kpc, resulting in a time unit $= 2.6$ Myr %and speed of light $= 795.0$ in model units.
\subsection{Gravitational Softening}
We use a gravitational (Plummer-) softening between all particles.
%$\varphi$-GPU does not include the regularization \citep{mikkola98} of close encounters or binaries, instead 
 $\varphi$-GPU supports the use of different softening lengths for different components and even individual softening for the particles.  
The softening between the SMBH particles is set equal to zero. But the use of zero softening for the stars and dark matter leads to the 
formation of tight binaries in the system, which causes an enormous slow down because of small time steps required to resolve the orbits of the
binaries. $\varphi$-GPU does not include the regularization \citep{mikkola98} of close encounters or binaries, so we use softening to avoid the formation of tight binaries. For the interactions between two particles, we adopt the following criteria for the gravitational softening: 

\begin{equation}
\epsilon_{ij} ^2 = (\epsilon_i ^2 + \epsilon_j ^2)/2 ,
\end{equation} \label{soft} 
 where $\epsilon_{bh} = 0$ for SMBHs, $\epsilon_{s} = 0.01$ pc for stars and $\epsilon_{dm} = 10$ pc for dark matter particles.

During the late phase of the SMBH binary evolution (Run-3), we reduce the star-BH interaction softening additionally by a factor of ten to resolve the stellar encounters during 
the pericenter passage of the SMBH binary.
%\begin{itemize}
%\item To calculate the forces between dark matter particles and stars, the softening $\epsilon_{ds} = \sqrt{(\epsilon_d^2 + \epsilon_s^2)/2}$ is used.
%\item To calculate the forces between dark matter particles and the SMBHs, the softening $\epsilon_{db} = \sqrt{(\epsilon_d^2 + \epsilon_b^2)/2}$ is used.
%\item To calculate the forces between the SMBHs and star particles, the softening $\epsilon_{bs} = 0.1\cdot\sqrt{(\epsilon_b^2 + \epsilon_s^2)/2}$ is used.
%\end{itemize}
 
%  A softening of 0.001 pc is necessary to bring the SMBH binary separation to milli parsec regime where GW emission become efficient. 

\subsection{Time Step Criterion}
The applied time step criterion \citep{nk08} for individual particles is 

\begin{equation}
\Delta t = \eta \left(\frac{A^{(1)}}{A^{(p-2)}} \right) ^{1/(p-3)} ,
\end{equation} \label{TC1}

where

\begin{equation}
A^{(k)} = \left(|a^{(k-1)}||a^{(k+1)}| + |a^{(k)}|^2 \right)^{1/2}.
\end{equation} \label{TC2}

Here $p$ is the order of the integrator, $a^{(k)}$ is the k-th order derivative of the acceleration and $\eta$ is the time step parameter.

 $\varphi$-GPU can employ different time step parameters $\eta$ for different components (BH, stars and dark matter). We adopted $\eta = 0.1$ for all components. In the late phase of the SMBH binary evolution (Run-4), we reduce this parameter for SMBHs, $\eta = 0.3$, to integrate the binary orbit with smaller time steps, hence achieving higher accuracy. 

\subsection{GPU Clusters}
The $N$-body integrations were carried out on three GPU high-performance computing clusters. 
{\tt laohu} employing 172 GPUs at the Center of Information and Computing at National Astronomical Observatories, Chinese Academy of Sciences. {\tt titan} having 32 GPUs at the Astronomisches Rechen-Institut in Heidelberg, and {\tt accre} employing 192 GPUs at Vanderbilt University, Nashville, TN. 

We used up to 64 GPUs for our runs with a total CPU wall-clock time of approximately one year. 

\section{SMBH Binary Evolution}

We start our high resolution run at time ($T=0$), when the two SMBHs are still embedded in their respective galaxy cusps. Figure~\ref{fig1zu} shows the dark matter (top) and stellar volume mass densities (bottom) with view on the $x-y$ (left) and $y-z$ (right) plane. For visualization we use the open-source $\tt glnemo2$ software package\footnote{\tt 
http://projets.oamp.fr/projects/glnemo2}. Two high density regions around the two SMBHs are clearly visible in the figure which suggest that individual galaxy cusps are still in the process of merging at the beginning of our run. The evolution of the SMBH binary can be described in the following three distinct phases as discussed in the following subsections.

\subsection{Dynamical Friction} 
In the first phase, the two black holes centered in their respective galaxy cusps are unbound to each other and move independently in the potential of the merger remnant. Dynamical friction against the background dark matter and stars is very efficient in bringing the two SMBHs closer. At about $T = 40$ Myr the individual cusps are already merged into one and the two SMBHs are located in a single cusp (see Figure \ref{fig3}). Earlier studies show that SMBHs form a binary when their relative separation $\Delta R_{BH}$ reaches $\sim r_h$ ($r_h$ is the gravitational influence radius defined as the radius of a sphere around the two black holes enclosing stellar mass equal to twice the SMBHs masses). 
%The combined mass of the two black holes in our simulations is $6.0 \times 10^{-5}$~ (in model units). 
Figure \ref{fig4} shows the evolution of the binary separation. At the same time when the two cusps merge ($T = 40$ Myr) the separation between the two SMBHs is roughly about $r_h = 15 ~$pc and a SMBH binary system is formed. 

%\item In the second phase, the binary separation shrinks very rapidly due to the combined effect of dynamical friction and gravitational slingshot effect which efficiently draw energy %and angular momentum away from the SMBH binary. Judging from figure \ref{fig4}, this phase of rapid binary evolution ends somewhere around $T = 110-120$~ Myr. The contribution from %these two mechanisms (i.e .dynamical friction and gravitational slingshot effects) in this phase is difficult to disentangle however equations (13) and (14) of \citet{mil01} give %approximate estimates of energy transfer from the binary to the stars by these mechanisms. The motion of SMBH binary in this phase is approximately Keplarian.\footnote{The semi-major %axis $a$ and eccentricity $e$ of the binary are defined here via the standard Keplerian relations, i.e., neglecting effects of the field
% stars.} 
 
%\item Once the binary semi-major axis $a \approx a_h$, where $a_h$ is semi-major axis of the hard binary defined by,

% \begin{equation}
%a_{h} = \frac{q}{(1+q)^{2}}\frac{r_{h}}{4}
%\end{equation} \label{ah}
%\citep[e.g.][] {mer07}
%the rapid phase of the SMBH binary becomes to an end. 

%For this case as we noticed that $r_{h}$ is 15 pc , $a_{h} \approx 0.3$ pc with $a_{h}^{-1} \approx 3.5$ ~pc. From figure \ref{fig6}, we see that indeed, the rapid hardening comes to %an end when inverse semi-major axis of teh massive binary is approximately 4 pc.

\subsection{Stellar-Dynamical Hardening}

\begin{figure}
\centering
\includegraphics[angle=00, width=0.99\columnwidth]{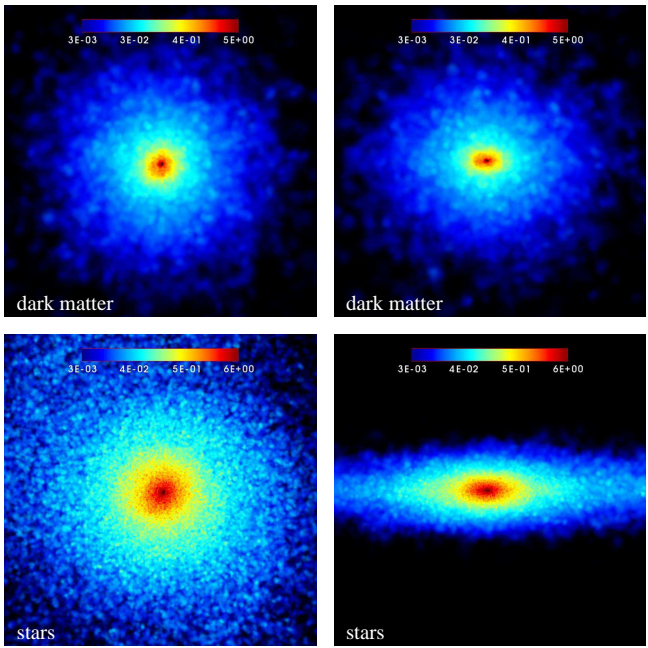}
\caption[]{
Same as in Figure 1 but at $T = 40$ Myr. Both SMBHs are embedded in a single cusp. 
} \label{fig3}
\end{figure}

\begin{figure}
\centerline{
  \resizebox{0.99\hsize}{!}{\includegraphics[angle=270]{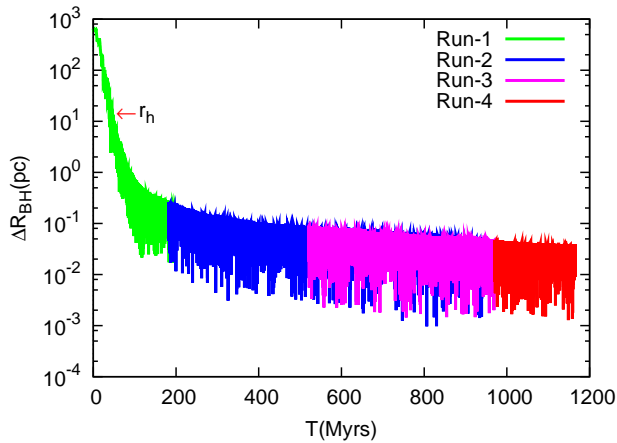}}
  }
\caption[]{
Relative separation between the two SMBHs as a function of time. The red arrow shows the estimated value of the influence radius $r_h$.
} \label{fig4}
\end{figure}  

The subsequent evolution of the binary is governed by stellar encounters, i.e., predominantly three body encounters. For spherical galaxy models the subsequent hardening is reported to depend on the particle number \citep{mak04,ber05}. For a realistic particle number $N$, which is several orders of magnitude larger than current state-of-the-art simulations can accommodate, the binary should stall when its semi-major axis $a \sim a_{h}$ for these models. Here $a_h$ is the semi-major axis of a ``hard binary", as defined by

\begin{equation}
a_\mathrm{h} = \frac{q}{(1+q)^{2}}\frac{r_\mathrm{h}}{4}
\end{equation} \label{ah}
\citep[e. g.][] {mer07}.

In our model $r_\mathrm{h}$ is about 15 pc, $a_\mathrm{h} \approx 0.66$ pc with $a_\mathrm{h}^{-1} \approx 1.5$ pc$^{-1}$.

\begin{figure}
\centering
\includegraphics[angle=270, width=0.99\columnwidth]{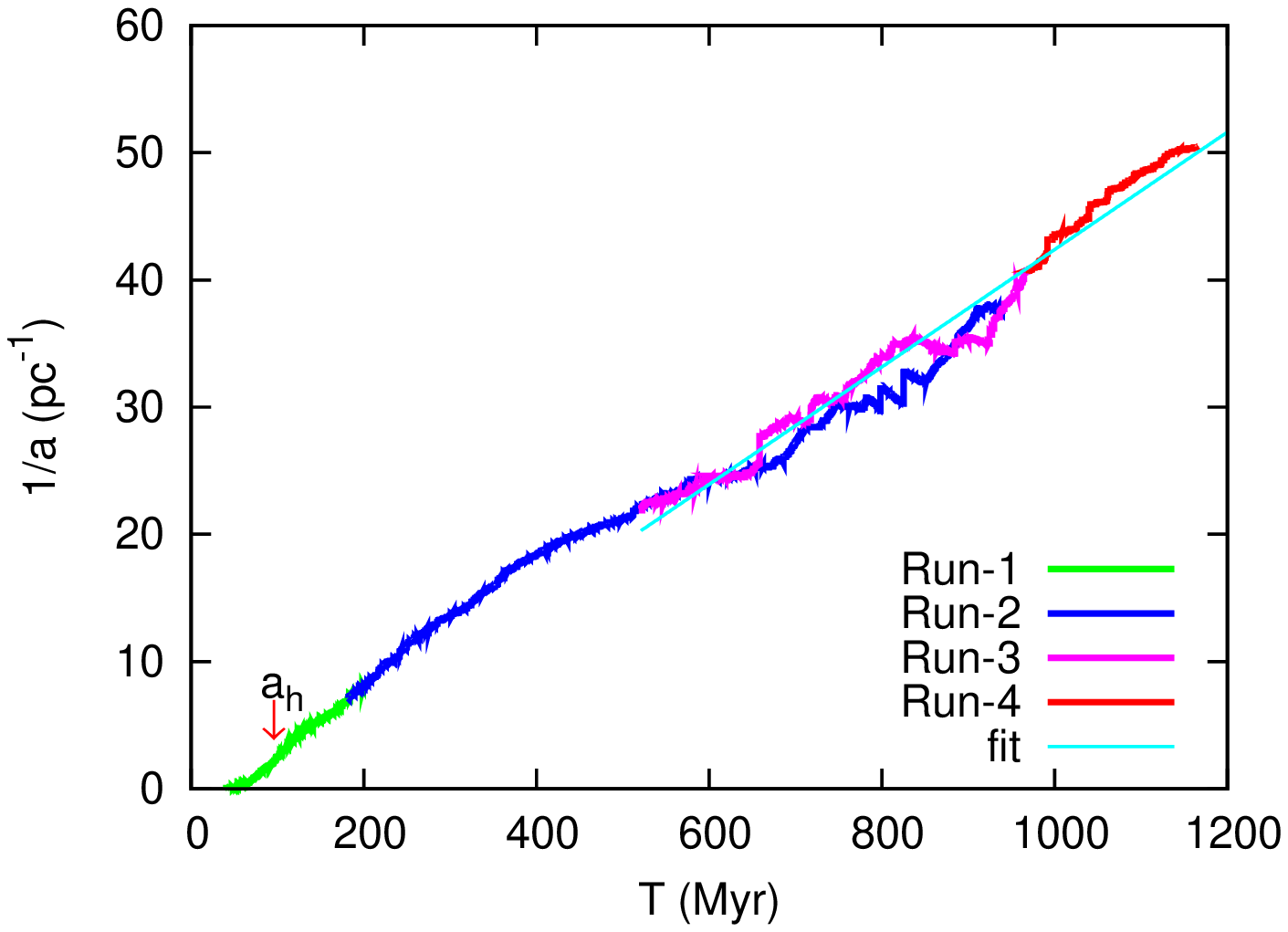}
\includegraphics[angle=270, width=0.99\columnwidth]{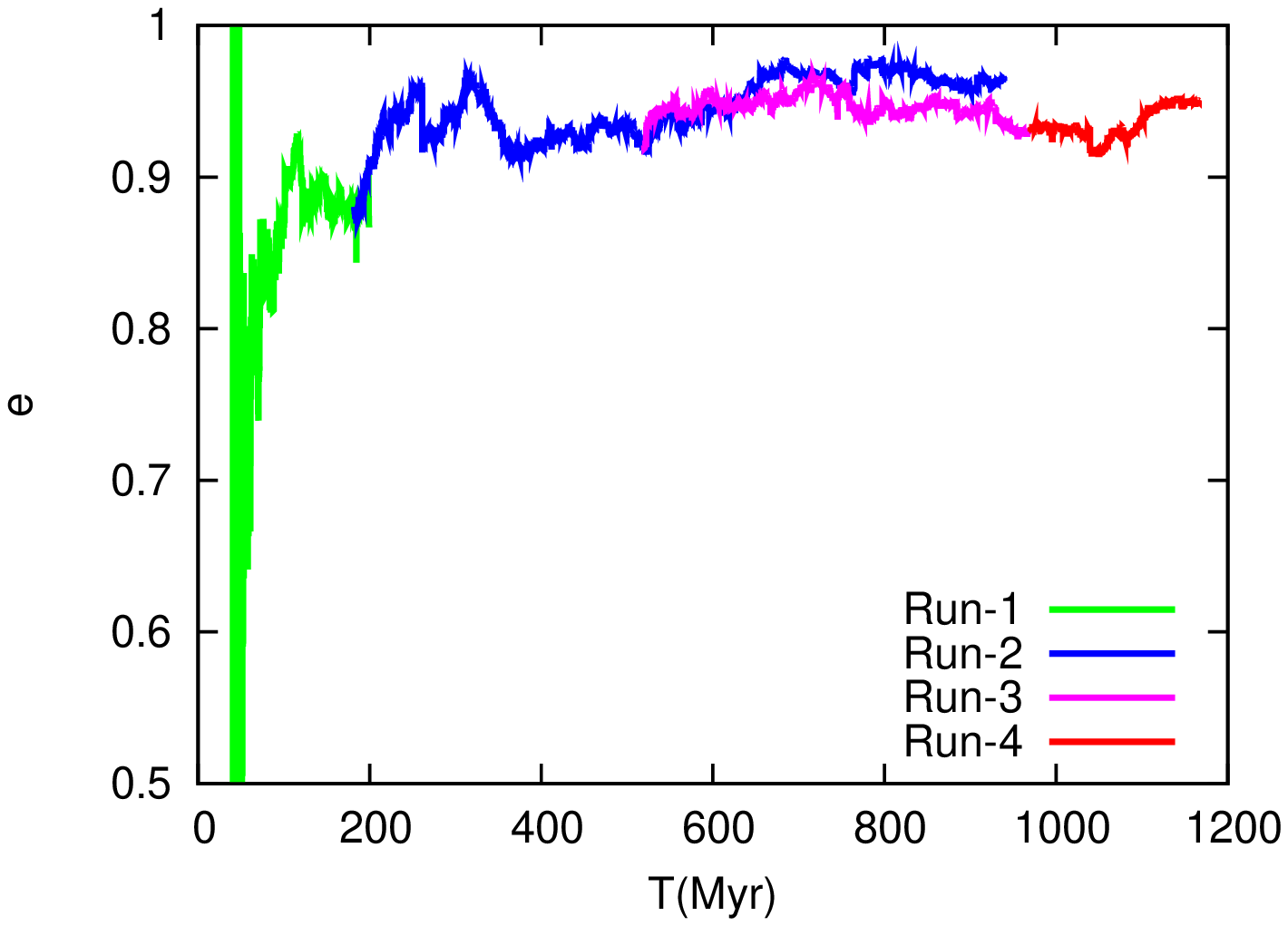}
\caption[]{
The evolution of the binary's inverse semi-major axis (top) and eccentricity (bottom). The red arrow in the upper panel of the figure points to value of $1/a$, which corresponds to the estimated semi-major axis $a_\mathrm{h}$ of the hard binary. The thin blue line shows the linear fit to estimate the constant SMBH binary hardening rate.
} \label{fig5}
\end{figure}

%However, it has been shown in previous studies, e.g., \citep{ber06, ber09, kha11,pre11} (PLEASE APPRECIATE EARLIER WORK, ESPECIALLY IF FROM SAME GROUP) that more realistic models of %merger remnants (which break the assumption of spherical symmetry) this $N$-dependence disappears. The non spherical shapes of merger remnant support a large fraction of stars on %centrophilic orbits \citep[e.g.][]{st99,pm01,MP04,mv11} which efficiently drive the hardening of SMBH binary. This suggests that result obtained in galaxy merger simulations can be %extrapolated to galaxy models with arbitrary large number of stars. 
Figure \ref{fig5} shows the evolution of the SMBH binary's inverse semi-major axis and eccentricity. The eccentricity grows to a very high value of $e \approx 0.9$ soon after the formation of the SMBH binary. The inverse semi-major axis of the binary evolution is roughly constant in time and the phase of ``hard binary" evolution is reached quickly. This behavior is consistent with the findings of \citet{kha12}, who noticed that for shallow cusps (with $\gamma$ = 0.5, 1.0) $1/a$ of the binary evolves at a constant rate immediately after its formation (see Figure 2 of \citet{kha12}). In fact, we find for stellar density profile, $\gamma = 1 =$ const in our simulation. For steep cusps ($\gamma$ = 1.5, 1.75) the SMBH binary undergoes a rapid phase of evolution before entering the hard binary regime, presumably corresponding to the clearing of the loss cone \citep{Yu02}. The long term evolution of the SMBH binary is discussed later. Here we describe in detail the different runs that we carried out in order to reach a SMBH separation where gravitational wave emission starts becoming important.
%\footnote{The loss cone is the region of phase space corresponding, roughly speaking, to orbits that cross the binary, i.e. with angular 
%momentum $J \lesssim J_\mathrm{lc} = \sqrt{GM_{\bullet}fa_\mathrm{bin}}$, where $f={\mathcal O}(1)$ \citep{1977ApJ...211..244L}.}

\begin{figure}
\centerline{
  \resizebox{0.99\hsize}{!}{\includegraphics[angle=270]{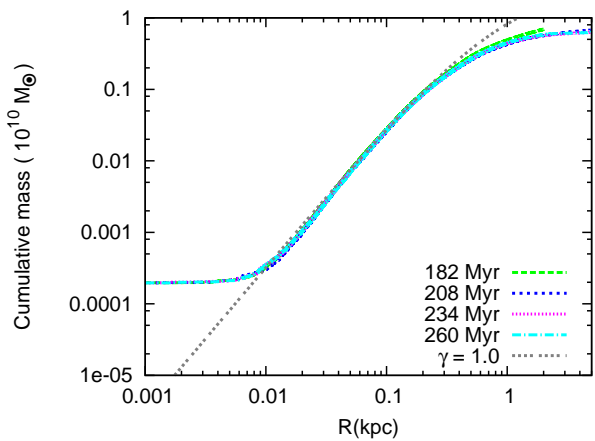}}
  }
\caption[]{
Mass profile after the selection of the new particle sample at $T = 182$ Myrs. The black dotted line represents theoretical Hernquist model $\gamma = 1.0$. Clearly the cumulative mass profile is very stable in the inner kpc.
} \label{fig6}
\end{figure}

At $T = 200$ Myr, we reduce the particle number from $\sim 3$ million to $\sim 1.15$ million by selecting the particles which have their estimated pericenter in the inner 1 kpc to increase the computational speed (Run-2). In order to see whether or not our selection has introduced some changes in the mass distribution, we plot the cumulative mass distribution at various time steps after the new selection of particles (Figure \ref{fig6}). The cumulative mass profile looks very stable in the inner parts. Only in the outer parts there is a small expansion of the profile as expected due to the new cutoff. Also we start the new run (Run-2) 20 Myr earlier to see, if the evolution of the binary as it happened in the earlier run (Run-1) can be recovered. Figure \ref{fig5} shows that both the inverse semi-major axis and the eccentricity evolution are well reproduced for the period where the two runs overlap.

\begin{figure}
\centering
\includegraphics[angle=270, width=0.99\columnwidth]{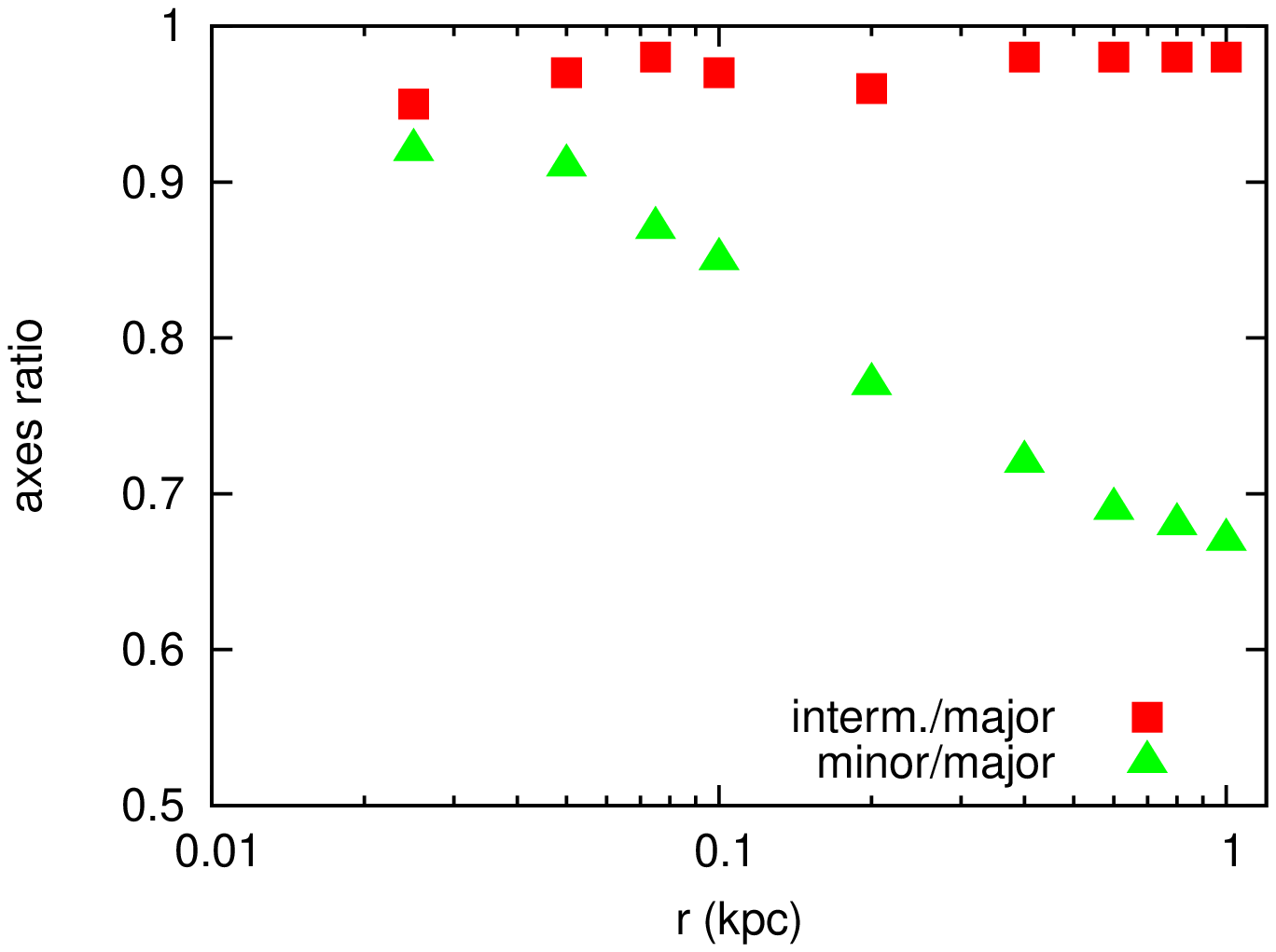}
\includegraphics[angle=270, width=0.99\columnwidth]{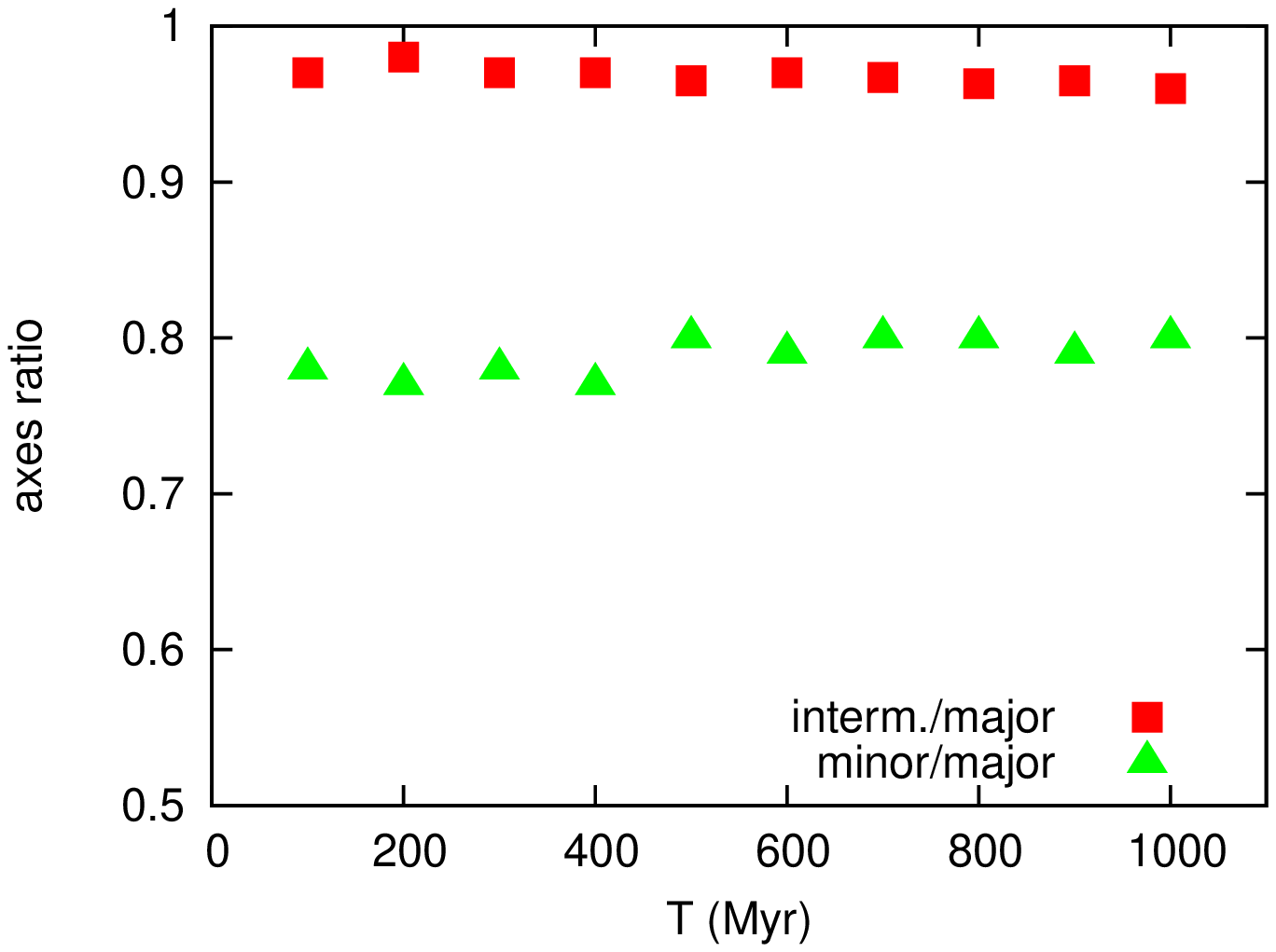}
\caption[]{
\textit{Top}: Intermediate to major  and minor to major axes ratios as a function of distance from the center of the SMBH binary at $T = 200$~Myr. \textit{Bottom}: The time 
evolution of axes ratios calculated at a distance of $0.2$ kpc from the center of the SMBH binary.  
} \label{fig7}
\end{figure}

We stop Run-2 when the value of the inverse semi-major axis is roughly 40 pc$^{-1}$ or $a = 25$ mpc. The value of eccentricity at the end of Run-2 is $e \sim 0.96$. For these parameters the value of the pericenter r$_\mathrm{p}$ for the massive binary is $r_\mathrm{p} = (1-e)a = 1$ mpc, which is smaller than the softening for the star-BH interaction (7 mpc). In order to resolve the star-BH encounters accurately at the pericenter passage of the binary, we reduced the softening for star-BH encounters by an additional factor 10 and start a new run (Run-3) at an earlier time of 520 Myr. For Run-3, we again split the dark matter particles, this time only those having pericenter smaller than 50 pc from the center of the massive binary.  As for the earlier splitting, each dark matter particle is split into ten particles and spread over a 10 pc sphere retaining the velocities of the parent particles. We employ the new particle splitting to avoid the unphysical jumps in the binary semi-major axis caused by massive dark matter particles that occur from time to time. Again, our new particle splitting does not introduce noticeable changes in the central mass profile. 
For Run-4, we reduce the $\eta$ parameter for the SMBHs from $\eta = 0.1$ to $\eta = 0.1$ to achieve higher accuracy in the integration of the SMBH binary orbit (there are roughly $10^4$ orbits in one model time unit).  We evolve the SMBH binary till about 1.2 Gyrs with Run-4. The inverse semi-major axis value is 50 pc$^{-1}$ or $a = 20$ mpc. The eccentricity value is 0.955, which leads to the pericenter distance of 0.9 mpc. The binary evolution for Run-2 is very similar to Run-3 and Run-4. The binary's inverse semi-major axis evolves at a constant rate (top panel of Figure \ref{fig5}), which is consistent with our earlier studies where we followed the evolution of SMBH binary by merging two spherical galaxies \citep{kha11,kha12}. We fit a straight line to calculate the binary's hardening rate $s = \frac{d}{dt}(1/a)$ in the late phase of Run-3 and Run-4 (Figure \ref{fig5} - top panel). The value of the hardening rate is 115.7 in model units and 44.5 kpc$^{-1}$ Myr$^{-1}$ in physical units.
This value of the hardening rate is similar to those obtained by merging two spherical galaxies (see top panel of Figure 8 of \citet{kha11}) having a similar profile ($\gamma = 1$) as adopted for the galaxy bulges in the merger study of \citet{cal11}. In \citet{kha11}, the high value of the hardening rate was attributed to the non spherical shape of the merger remnant supporting a large fraction of stars on centrophilic orbits (see also \citet{ber06}). The value of the hardening rate is approximately 6 times higher for the same $N$ when compared to the value for similar mass of SMBHs in spherical galaxy models (top panel of Figure 3 of \citet{kha11}).
For the merger remnant of two late type galaxies under consideration in our current study, we analyze the shape by calculating axes ratios defined for a homogeneous ellipsoid with same tensor of inertia. Figure \ref{fig7} shows the intermediate to major and minor to major axes ratios at various distances from the center (top panel) and also for different times (bottom panel). We can see from the top panel of the Figure \ref{fig7} that deviations from spherical symmetry extend all the way to the center (few tens of parsec). The merger remnant is considerably flattened, which can also be seen from Figure \ref{fig3}, when compared to those which result after the merger of two spherical galaxy progenitors.  The flattening increases as we move to larger distances and becomes more or less constant about a distance of 1 kpc. We also calculate the axes ratio at different times of evolution for the merger remnant at a distance of 200 pc from the center, as most of the centrophilic orbits are expected to come at about this distance. As is seen from the bottom panel of Figure \ref{fig7}, the axes ratios remain constant during the whole time of the evolution of the SMBH binary.  Due to the non-spherical shape of the merger remnant, we expect that the SMBH binary should evolve at a constant rate supported by the centrophilic orbit family of stars rather than the relaxation effects alone. 
Therefore it is reasonable to extrapolate our results for the merger of late-type galaxies to a realistic number of star particles. Hence we can predict the coalescence time for the SMBHs using the estimated hardening rates in the stellar dynamical phase plus those in the GWs dominated regime.  

\begin{figure}
\centering
\includegraphics[angle=270, width=0.99\columnwidth]{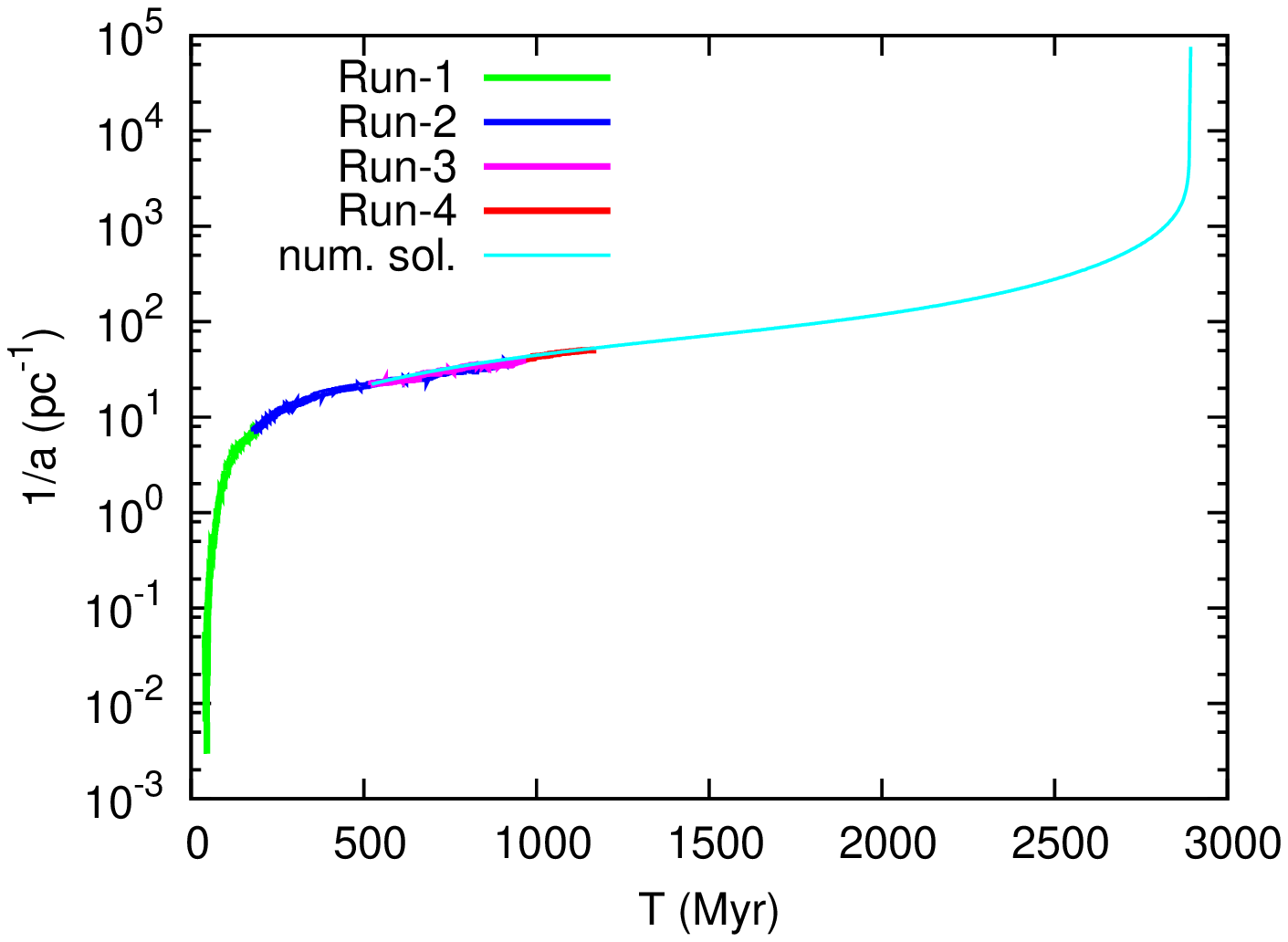}
\includegraphics[angle=270, width=0.99\columnwidth]{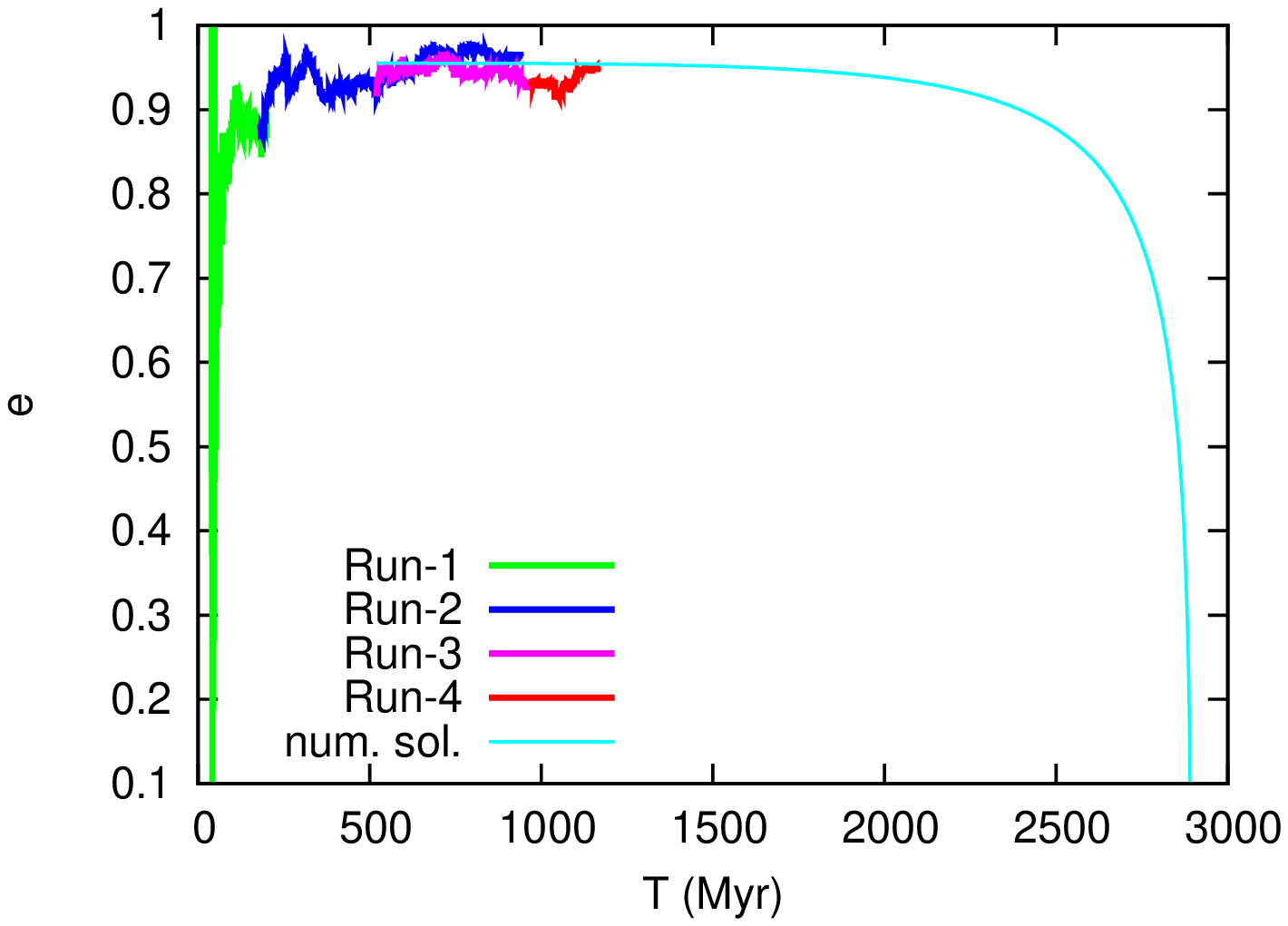}
\caption[]{
The evolution of the binary's inverse semi-major axis (top) and eccentricity (bottom). Run1 to Run4 describe the evolution of the SMBH binary during direct $N$-body simulations performed using $\varphi$-GPU. Thin blue line is the numerical solution of the coupled equations (\ref{ratea}) to (\ref{dedt}) for the estimate of the SMBH binary evolution.  
} \label{fig8}
\end{figure} 
%\section{Time for the Coalescence of SMBH Binary} \label{coalzu}

% We chose this approach, because more self-consistent 
%$N$-body simulations including relativistic effects for the SMBH binary \citep[see][]{berent09}, 
%are computationally expensive. \cite{berent09} have shown that in GWs dominated regime Peters 
%formula accurately predicts the coalescence time of the SMBH binary.  

\subsection{Relativistic Regime}
 At small enough separation whose value depends on the mass and eccentricity of the SMBH binary, gravitational waves extract energy and angular momentum efficiently from the binary, thus making its coalescence inevitable.
%It is worth to mention that the LISA will be most sensitive for the SMBHs having masses comparable to those which are adressed in this study. The SMBH with such masses ($10^5 - 10^6 %M_{\odot}$) are observed in centers of late type galaxies. So in this study, we follow the evolution of massive binary in this interesting mass regime starting from consisten intial %set up of merging galaxies. We present the SMBH coalescence time estimate in the next section.  

As is shown in earlier studies \citep{ber09,kha12}, the estimated coalescence time obtained using constant hardening rate $s$ in the stellar dynamical regime and the formula of \citet{pet64} for hardening in the gravitational wave dominated regime agree remarkably well with $T_{coal}$ obtained from simulations that follow the binary evolution till coalescence using post-Newtonian ($\mathcal{PN}$) terms in the equation of motion of the SMBH binary. But these simulations are computationally very expensive due to additional $\mathcal{PN}$ terms in the equation of motion of the binary and small softening needed to resolve the star-BH interactions at pericenter till the SMBH binary enters in the gravitational wave dominated regime. In order to further evolve the binary to the full coalescence of the SMBHs, we need to again reduce the softening and also add $\mathcal{PN}$ terms in the equation of motion of the binary, which would increases the computational time drastically (by several months).

Further evolution of the SMBH binary can be estimated by

\begin{equation}
\frac{da}{dt} = \left(\frac{da}{dt}\right)_\mathrm{NB} + \Big\langle\frac{da}{dt}\Big\rangle_\mathrm{GW} = -s a^{2}(t) + \Big\langle\frac{da}{dt}\Big\rangle_\mathrm{GW} \label{ratea}
\end{equation}

The orbit-averaged expressions---including the lowest order 2.5 $\mathcal{PN}$ dissipative terms--- for 
the rates of change of a binary's semi-major axis, and eccentricity due to GW emission are
given by \citet{pet64}:
\begin{mathletters}
\begin{eqnarray}
\Big\langle\frac{da}{dt}\Big\rangle_\mathrm{GW} &=& -\frac{64}{5}\frac{G^{3}M_{\bullet1}M_{\bullet2}(M_{\bullet1}+M_{\bullet2})}{a^{3}c^{5}(1-e^{2})^{7/2}}\times \nonumber \\
&&\left( 1+\frac{73}{24}e^{2}+\frac{37}{96}e^{4}\right),  \label{dadt}\\
\Big\langle\frac{de}{dt}\Big\rangle_\mathrm{GW}  &=& -\frac{304}{15}e\frac{G^{3}M_{\bullet1}M_{\bullet2}(M_{\bullet1}+M_{\bullet2})}{a^{4}c^{5}(1-e^{2})^{5/2}}
\times\nonumber\\
&&\left( 1+\frac{121}{304}e^{2}\right) .  \label{dedt}
\end{eqnarray}
\end{mathletters}

Our estimates show that already for the parameters of the SMBH binary at $T \sim 1.2$ Gyr, the contribution to the hardening of the SMBH binary can be as large as 10\%. This is the reason that we stop our Run-4 at this point.

We now solve the coupled equations (\ref{ratea}) to (\ref{dedt}) numerically to follow the SMBH binary evolution. For a numerical solution of the coupled equations, the semi-major axis of the binary was chosen at a time $T \sim 500$ Myr to have a significant overlap with the $N$-body evolution of the massive binary. The eccentricity value was chosen to be 0.95 and we assume that the eccentricity remains constant during the stellar dynamical hardening phase. This assumption is supported by the eccentricity evolution shown in Figure \ref{fig5} (bottom) which shows that the value of e remains more or less constant from time $T \sim 600$ Myr onwards. 

The estimated evolution is shown in Figure \ref{fig8}. We can see that the coalescence time of the SMBH binary is $T_{coal} \sim 2.9$ Gyrs. The coalescence time of 2.9 Gyr, though longer when compared to \citet{kha11,pre11,kha12}, is still short enough to have a handful 1:10 merger cases for the detection with eLISA. From \citet{kha12}, we know that binary hardening rates depend strongly on the adopted density profile. For steep density cusps with an inner power law density index $\gamma = 1.75$, their study shows 4 times higher values of $s$ when compared to $\gamma = 1.0$. In the current study the adopted density profile at the start of the merger simulation was a Hernquist profile, which has $\gamma =1$. This slope is observed in bright elliptical galaxies which host SMBHs having masses $\sim 10^8-10^9$ M$_\odot$. The faint bulges/ellipticals which host smaller SMBHs 
with masses $\sim 10^6-10^7$ M$_\odot$ have typically steep cusps ($\gamma \sim 1.5-1.75$).  
It is conceivable that the prolonged effect of gas dissipation at higher mass and spatial resolution in the last stages of the merger, 
beyond the starting point of the direct $N$-body calculation, 
would have led to a steeper baryonic cusp at small scales (see Mayer et al. 2007 on the dependence of the inner density profile of merger remnants on 
the numerical resolution of the gas component). 
In addition, the slope of the initial bulge profile in the galaxy models could be steeper and yet still consistent with the observed distribution of bulge
slopes. 
%(see e. g. Gadotti et al. 2008).
Hence,  we can easily expect that typical coalescence times for SMBH binaries in gas-rich mergers can be shorter and comparable to \cite{kha12}.

\section{Summary \& Conclusions}

Starting from the results of \citet{cal11}, we studied the orbital evolution of a pair of SMBHs in a minor merger of disk galaxies (with 30\% gas fraction 
in the disk), from an initial separation of 60 kpc to a final separation of less than a milli parsec (binary's pericenter distance). Initially the mass ratio between the galaxies and SMBHs is 0.1. During the merger the two SMBHs accrete gas and increase their masses in the process. The mass of the SMBH in the satellite galaxy increases almost 8 fold as the gas in the secondary galaxy is funneled towards the center due to the tidal force of the primary galaxy at each peri-center passage. The perturbations produced by the 
passages of the secondary galaxy are not significant for the primary galaxy, so the SMBH in the primary galaxy accretes gas steadily and the mass of SMBH here grows by a factor of $2$. 
At the end of SPH simulations, the mass ratio between the two SMBHs is approximately $q=0.3$ (see Figure 1 of \citet{cal11}).

At the start of our direct $N$-body simulations, the separation between the two SMBHs is roughly 700 pc, and the binary has yet to form. 
Gas particles, which contribute only a few percent to the mass of the selected central region, are treated as star particles. We use particle splitting to reduce the 
mass of dark matter particles to avoid both mass segregation and unphysical encounters of high mass dark matter particles with the SMBHs. Dynamical friction is very efficient in bringing 
the two SMBHs to a separation where they form a binary at a separation of roughly 15 pc.
The subsequent hardening, which happens at a constant rate, is governed by individual stars interacting with the massive binary. We artificially 
suppress the contribution of dark matter to the hardening of the SMBH binary by introducing a large softening ($\epsilon_\mathrm{dm} = 10$~pc). 
%Although we split the dark matter particles, the 
%mass of a dark matter particle is only a factor 10 smaller than the mass of the SMBH in the secondary galaxy. Thus the large softening is used to obtain a smooth evolution of the 
%binary. 
%{\bf Lucio: I commented out the above sentences because they sound too technical for the conclusions, and they have already been explained in the previous sections}.
The shape analysis of the merger product shows that the system is predominantly
triaxial from the periphery to the center. The SMBH binary evolves at a constant rate and the hardening rate 
is high, which suggests that the stalling of the SMBH binary should not be an issue in realistic galaxy 
mergers such as those considered here. The eccentricity is very high ($e \sim 0.95$) as was observed for the shallow density profile  ($\gamma \leq 1.0$) galaxy merger simulations 
performed in our earlier studies \citep{kha11, kha12}. The dependence on 
eccentricity of the coalescence time under GW emission is $T_\mathrm{coal,GW} \sim (1-e^2)^{7/2}$. For very eccentric SMBH binaries this could easily
account for a decrease of an order of magnitude. Accordingly it will be very important to further investigate the dependence of the eccentricity evolution under
different values of $\gamma$ and $q$. Currently we are carrying out direct $N$-body simulations of galaxy mergers with galaxies having both steep and shallow density profiles at the centers together with an initial mass function to address this question.    

With our current study we evolve the SMBH binary to a separation (0.9 mpc at pericenter of the SMBH binary), where the contribution to the hardening rate of the SMBH binary due to the emission of GWs becomes important (roughly 10\%). Using the constant value of the hardening rate in the Newtonian regime and the formula of \citet{pet64} for GWs emission from two point masses orbiting each other, in the relativistic regime, we estimate the coalescence time of two SMBHs to be 2.9 Gyr after the merger of the two galaxies. The coalescence time of 2.9 Gyr, although longer when compared to the times obtained for similar mass binaries in \citet{kha12}, is still short enough to have a few 1:10 mergers 
of SMBHs in late type galaxy mergers in the range at which eLISA is most sensitive. From \citet{kha12}, we know that binary hardening rates depend strongly on the adopted density 
profile. 
%{\bf Lucio: the paragraph below on the slope effect should be shortened into one single sentence since we have explained everything in the previous section}.
For steep density cusps observed at the center of faint bulges/ellipticals, we can expect the coalescence times to be much shorter, comparable to the ones that were obtained in \citet{kha12} for the merger of steep power law density profile galaxies.

%For steep density cusps having an inner power law density index $\gamma = 1.75$, the study shows a factor of 4-5 higher value of $s$ when compared to $\gamma = 1.0$. In the 
%current study the adopted density profile at the start of the merger simulation was a Hernquist profile, which has $\gamma =1.0$. This slope is observed in bright elliptical galaxies 
%which host SMBHs having masses $\sim 10^8-10^9$ M$\odot$. The faint bulges/ellipticals which host smaller SMBHs with masses $\sim 10^6-10^7$ M$\odot$ typically have steep 
%cusps ($\gamma \sim 1.5-1.75$).  So for an appropriate density profile, we can expect the coalescence times to be much shorter, comparable to the ones that were obtained in %\citet{kha12} for the merger of steep power law density profile galaxies. 
The current work should be regarded mainly as proof-of-concept, since we have considered
only one particular initial condition and have not computed directly the binary SMBH shrinking
in the post-Newtonian phase. Nevertheless,
 it shows for the first time that the coalescence
of the two SMBHs on a timescale sufficiently short to be astrophysically relevant 
does indeed take place as a result of a quite realistic galaxy merger with previous
effects of dissipation taken into account. In the future we will explore a wider range
of initial conditions motivated by cosmological simulations of galaxy formation and we 
will carry out the direct computation of the binary shrinking to much smaller
separations.

\acknowledgments

FK was supported by a grant from the Higher Education Commission (HEC) of 
Pakistan administrated by the Deutscher Akademischer Austauschdienst (DAAD).
IB and PB acknowledge financial support by the Deutsche Forschungsgemeinschaft (DFG) through 
SFB 881 ``The Milky Way System" (sub-projects A1 and Z2) at the Ruprecht-Karls-Universit\"at Heidelberg.

PB acknowledges the special support by the NASU under the Main Astronomical Observatory 
GRID/GPU computing
cluster project and by the program Cosmomicrophysics of NASU. 

We acknowledge Jean-Charles Lambert for his help and support regarding
 the visualisation program {\tt glnemo2} which has been used
 to produce some figures in this publication.

A main part of the simulations presented here was performed on the GPU cluster {\tt laohu} at
the Center of Information and Computing at the National Astronomical Observatories, Chinese
Academy of Sciences, funded by the Ministry of Finance of
People's Republic of China under the grant ZDYZ2008-2.

We also used the GRAPE/GPU cluster {\tt titan} funded through the GRACE project under the grants I/80 041-043
and I/81 396 by the Volkswagen Foundation and under the grants 823.219-439/30 and /36 of the Ministry of
Science, Research and the Arts of  Baden-W\"urttemberg, Germany. Finally, part of this work was conducted using the
resources of the Advanced Computing Center for Research and Education at Vanderbilt University,
Nashville, TN, USA.

---------------------


%% in parentheses, after the keyword, but they will not be verified.

%%{\it Facilities:} \facility{Nickel}, \facility{HST (STIS)}, \facility{CXO (ASIS)}.



\clearpage

%% the file, uncomment the \plotone, \plottwo, and
%% If you need a layout that cannot be achieved with \plotone or
%% \plottwo, you can invoke the graphicx package directly with the
%% \includegraphics command or use \plotfiddle. For more information,
%% please see the tutorial on "Using Electronic Art with AASTeX" in the
%% documentation section at the AASTeX Web site,
%% http://www.journals.uchicago.edu/AAS/AASTeX.
%%
%% The examples below also include sample markup for submission of
%% supplemental electronic materials. As always, be sure to check
%% the instructions to authors for the journal you are submitting to
%% for specific submissions guidelines as they vary from

%% This example uses \plotone to include an EPS file scaled to
%% 80% of its natural size with \epsscale. Its caption
%% has been written to indicate that additional figure parts will be
%% available in the electronic journal.

%%\begin{figure}
%%\epsscale{.80}
%%\plotone{f1.eps}
%%\caption{Derived spectra for 3C138 \citep[see][]{heiles03}. Plots for all sources are available
%%in the electronic edition of {\it The Astrophysical Journal}.\label{fig1}}
%%\end{figure}

%%\clearpage

%% Here we use \plottwo to present two versions of the same figure,
%% one in black and white for print the other in RGB color
%% for online presentation. Note that the caption indicates
%% that a color version of the figure will be available online.
%%

\end{document}